\documentclass[12pt]{article}
\newcommand{\beq}{\begin{equation}}
\newcommand{\eeq}{\end{equation}}
\newcommand{\beqa}{\begin{eqnarray*}}
\newcommand{\eeqa}{\end{eqnarray*}}
\newcommand{\beqna}{\begin{eqnarray}}
\newcommand{\eeqna}{\end{eqnarray}}
\newcommand{\beql}{\begin{equation}\begin{array}{l}}
\newcommand{\eeql}{\end{array}\end{equation}}
\newcommand{\n}{n_o}
\newcommand{\fr}{\frac}

\begin{document}
%\draft
%\baselineskip 25pt
%\pagestyle{plain}
%\bibliographystyle{plain}
%\interlinea{1.4}
%\setcounter{secnumdepth}{0}
%\baselineskip 25pt
%\title{
\begin{center}
\begin{Large}
A New Regime of Anomalous Penetration of Relativistically Strong Laser
Radiation into an Overdense Plasma\\
\end{Large}
%}
%\author{
A.~Kim,$^1$ F.~Cattani,$^2$ D.~Anderson,$^2$ and M.~Lisak$^2$\\
%}
%HO COMMENTATO ADDRESS CHE VA TENUTO IN REVTEX!!!!!
%\address{
$^1$Institute of Applied Physics, Russian Academy of Sciences, 603600
Nizhny Novgorod, Russia\\
$^2$Department of Electromagnetics, Chalmers University of Technology,
S-412 96 G\"{o}teborg, Sweden
%}
\end{center}
%commentare la data prima di mandare il file a jetp lett.

%\date{\today}

%\maketitle

\begin{abstract}
It is shown that penetration of relativistically intense laser light into
an overdense plasma, accessible by self-induced transparency, occurs over a
finite length only. The penetration length depends crucially on the
overdense plasma parameter and  increases with increasing incident intensity after
exceeding the threshold for self-induced transparency. Exact analytical
solutions describing the plasma-field distributions are presented.\\
PACS number(s): 52.40.Nk, 52.35.Mw, 52.60.+h, 52.58.Ns
\end{abstract}

%\pacs{PACS number(s): 52.40.Nk, 52.35.Mw, 52.60.+h, 52.58.Ns}

In the past few years there has been much research devoted to the nonlinear
interaction of superintense laser pulses with plasmas \cite{ref}. At
intensities where electrons quiver with relativistic velocities, the
interaction can be characterized as nonlinear optics in relativistic
plasmas, and new regimes, not evident at nonrelativistic intensities, may
appear. As was previously shown, superintense electromagnetic radiation
can propagate through a classically overdense plasma due to
the relativistic correction to the electron mass, the so called induced
transparency effect,
\cite{ap}-\cite{schep}.  The present work
has resulted in the identification of a new fundamental process in
relativistic laser overdense plasma interaction.

In order to understand the nonlinear regime of interaction of superintense
laser light with an overdense plasma, it is enough, without loss of
generality, to consider a stationary model. We present here a new class of exact
analytical solutions describing the penetration of an electromagnetic wave
normally incident onto a cold, overdense plasma with a sharp boundary.
In particular, we show that, when the incident intensity exceeds the
threshold for
self-induced transparency, the laser energy penetrates into the dense
plasma without any losses, but over a finite length only. At the same time,
the electron density distribution becomes structured as a sequence of
electron layers separated by about half a
wavelength wide depleted regions, so that this strongly nonlinear plasma
structure acts as a distributed Bragg reflector.

The ultrahigh intensity laser-plasma interaction is described by the
relativistic equation of motion and the equation of
continuity for the electrons together with Maxwell's equations. Ions are treated as a uniform
neutralizing background. We will consider circularly polarized laser
radiation with normalized amplitude of the vector potential
$e{\mathbf{A}}/mc^2 =(a/\sqrt{2})Re[({\mathbf{y}}+i{\mathbf{z}})exp(i\omega
t)]$
normally incident from vacuum ($x<0$) onto a semi-infinite plasma  ($x\ge
0$). Assuming a stationary regime, the basic equations may be written in
the form
%(see, for example, \cite{marburger}):
\begin{eqnarray}
& & \phi^{\prime\prime}=\n(n-1),\\
& & a^{\prime\prime}+(1-\frac{\n}{\gamma}n)a=0, \\
& & \phi^\prime=\gamma^\prime \;\; \textrm{if and only if } n(x)\ne 0.
\end{eqnarray}
(where variables are normalized as
$x\Rightarrow \omega x/c$, $n\Rightarrow n/n_o$,
$n_o=\omega_{p}^{2}/\omega^2$, $\omega$ is the carrier frequency of
the laser radiation, $\omega_{p}$ is the plasma frequency of the initial
unperturbed plasma, $\gamma=(1+a^2)^{1/2}$ is the relativistic factor, $n$
and $\phi$ are normalized electron density and scalar potential of the
plasma respectively). Eq. (3) indicates that only in the region where the electron
density $n(x) \ne 0$ the ponderomotive force, $\gamma^\prime (x)$, must
be compensated by the force of a longitudinal field. This statement
will be important for constructing solutions of interest in the present
analysis.

For homogeneous ion density, the system has the following Hamiltonian
\begin{equation} \label{int}
{\mathcal{H}}_E=\frac{1}{2 (1 + a^2)}
a^{\prime 2}-\frac{1}{2}(2 n_{o}\sqrt{1 + a^2}-a^2)
\end{equation}
which was analyzed in \cite{marburger}.

As we
are interested in a semi-infinite plasma, we first consider the case when
$n(x) \rightarrow 1$ and both $a(x)$ and $a^\prime (x)$  vanish as $x
\rightarrow \infty$. In this case the integral of motion equals
\beq \label{int0}
{\mathcal{H}}_E=-\n \equiv {\mathcal{H}}_0.
\eeq
Eq. (\ref{int}) can then easily be integrated to yield the following
single-parametric solitary solution
\beq \label{sol}
a(x)= \fr{A_m \cosh \left[|\varepsilon_0|^{1/2}(x-x^{(0)})\right]}{\n
\cosh^2 \left[|\varepsilon_0|^{1/2}(x-x^{(0)})\right]-|\varepsilon_0|}
\eeq
where $\varepsilon_0 =1-\n $ is the dielectric permittivity of the plasma
and the parameter $x^{(0)}$ defines the position of the maximum of the
function (\ref{sol}) which is given by $ A_m=2[\n (\n-1)]^{1/2}$.

For $n_{o} > 1.5$ this solution contains a region where the electron
density is negative, which is clearly unphysical. Requiring the
minimum electron density $n_m=1-4(\n-1)^2$ to be positive, i.e. $n_m\ge 0$,
we obtain a condition on the background plasma density and an argument in
favor of the depletion region: for $\n>1.5$ we have to take into account
only that part of the solution where the corresponding electron density is
positive, $n(x)\ge 0$.By using the part of the solution (\ref{sol}) corresponding to
a ponderomotive force that pushes electrons into the plasma and matching it
to the vacuum one, an exact expression for the intensity threshold of self-induced
transparency was found, \cite{fede}. It should be noted that the Hamiltonian
(\ref{int0}) corresponds to a zero energy flux, but there are solutions with non-zero
flux as well, \cite{ap,lai}. They can arise even in correspondence to incident amplitudes
smaller than the penetration threshold. However, for the realization of such  nonlinear hysteresis-like
solutions, \cite{schep,litvak}, preliminary modifications of the plasma must be
induced by extremely intense fields. For the description of the steady-state
of a problem involving laser radiation with turn-on shape it is natural to choose the
zero flux approach.\\

The left side of the function (\ref{sol}) gives rise to an unbalanced  ponderomotive
force pulling
electrons out of the plasma towards the incident wave and, at first glance,
it seems that the charge
quasineutrality condition cannot be satisfied. However, after a length of about half a
wavelength, these electrons will be stopped by the ponderomotive force
acting in the opposite direction. Thus, in the general case, we may expect
that the full plasma-field structure will consist of a sequence of
alternating depletion and nondepletion regions. This can be understood from
 Fig. 1, where the phase portraits described by Eqs (1)-(3)and by
 the corresponding equation for the depletion (vacuum)
region are presented.  For a half-space plasma the limiting case
corresponds to a  motion with infinite period along the separatrix
determined by Eq. (\ref{int0}), i.e., an exponentially  decreasing field
inside the overdense plasma. Going
backwards towards the initial vacuum-plasma boundary, before the last
semi-infinite electron layer there must exist a depletion region. Here
the amplitude of the field corresponds to a forward going  wave along the
incident direction with an intensity below the threshold value (corresponding to the
motion along the first circular trajectory in phase space, coming from the
vacuum Hamiltonian). Then, in front of this layer we have to put another
electron layer, where the solution for the field $a(x)$ follows from the
Hamiltonian (\ref{int}) with a magnitude ${\mathcal{H}} > -\n$
(corresponding to an oscillatory motion about zero). This construction is
repeated  until the initial plasma boundary is reached. At the boundary between each
depletion and nondepletion region, the solution for the field must satisfy
continuity conditions for both $a$ and its derivative $a^\prime
$. It is also clear that there exists a
family of stationary solutions that differ from each other by the number of
electron density layers and their shapes. When the incident amplitude is
increased, the number of layers will increase as well, as follows from the
phase portraits in Fig. 1.

In order to quantify the above discussion, we present a more rigorous
analytical description. Starting from inside the plasma region, the
solution for the field is an exponentially decreasing function of the
spatial coordinate, in fact a part of  the localized solution given by Eq. (\ref{sol}) with
${\mathcal{H}}_E={\mathcal{H}}_0$ for $x_0\le x<\infty$.  The point $x=x_0$ can be
determined self-consistently from the global solution and the boundary
conditions
\begin{equation} \label{cont0}
a(x=x_0)=a_0, \; a^\prime (x=x_0)=a^\prime_0
\end{equation}

The next region ($x_1 < x < x_0$) must be a depletion layer where $n(x)=0$.
The second boundary position of this depletion layer, $x_1$, must also be
determined self-consistently. The Hamiltonian here is the vacuum one and
reads

\begin{equation}\label{intvacuum}
{\mathcal{H}}_V=\frac{1}{2}(a^{\prime 2} + a^2)=\frac{1}{2}(a^{\prime
2}_{0} + a_{0}^2)\equiv{\mathcal{H}}_1
\end{equation}
where we have taken into account the boundary conditions for the field and
its first derivative. The solution of the field in this depletion region reads
\begin{equation}
a(x)=A_1 \cos(x + \psi_1).
\end{equation}
Using (\ref{cont0}) we have: $A_1=(a^{\prime 2}_0 + a_0^2)^{1/2}=(2{\mathcal{H}}_1)^{1/2}$,
$\psi_1=-\arctan(a^{\prime}_0/a_0)-x_0$.
The boundary position $x_{1}$can be  calculated by integrating Poisson's
equation within the interval $x_1 < x < x_0$:
\begin{equation}
\phi^\prime(x_0) - \phi^\prime(x_1)=-n_{o}(x_0 -x_1).
\end{equation}
which, together with  relation (3), leads to a transcendental equation for
$\xi =x_0 - x_1$:
\begin{equation}\label{trasc}
\xi=g(\xi)-g(0)
\end{equation}
where
\begin{equation}
g(\xi)=\frac{A_1^2\sin[2(\xi+\xi_1)]}{2n_{o}[1+A_1^2\cos^{2}(\xi+\xi_1)]^{1/2}},
\; \xi_1=\arctan\left(\frac{a^{\prime}_0}{a_0}\right)
\end{equation}

Since the left-hand side of Eq. (\ref{trasc}) is a linear function while
its right-hand side is a periodic function of $\xi$, this equation has a
non-trivial solution only if $g^\prime (\xi=-\xi_1)>1$, i.e.
\begin{equation} \label{gprime}
g^\prime (\xi=-\xi_1)=\frac{A_1^2}{n_{o}(1+A_1^2)^{1/2}}>1.
\end{equation}
However, if $n_{o}\le 1.5$, this condition is never satisfied, because its
maximum value reaches one at $n_{o}=1.5$. Consequently, for plasma
densities $n_{o}\le 1.5$ we conclude that there are no stationary regimes
of anomalous penetration: there can only exist dynamical solutions. For
$n_{o}>1.5$, Eq. (\ref{trasc}) always has non-trivial solutions, which can
be found numerically. Formally, Eq. (\ref{trasc}) admits several
roots  but we have to consider only those that
correspond to a positive electron density.

Having solved Eq.(\ref{trasc}) we know $a_1=a(x_1)$ and
$a^\prime_1=a^\prime(x_1)$. The next region must again be an electron
layer. Denoting this region as $x_{2}\le x\le x_{1}$, the Hamiltonian
(\ref{int}) will here be
\begin{equation} \label{H02}
{\mathcal{H}}_E={\mathcal{H}}_E(a_1, a^\prime_1)\equiv {\mathcal{H}}_2.
\end{equation}
% >-\n
with the corresponding field solution written in terms of two-parameter
elliptic functions for the field strength as
\begin{equation} \label{sol1}
a(x)=\left\{
\begin{array}{l}
\frac{2q {\rm cn}[\varepsilon_2^{1/2}(x-x^{(2)})]}{2+[(q^2+1)^{1/2}-1]{\rm
sn}^2[\varepsilon_2^{1/2}(x-x^{(2)})]},\: {\mathcal{H}}_2<\n, \\
\frac{2\bar{q}{\rm
sn}[((\varepsilon_2+1)^2-\n^2)^{1/2}(x-x^{(2)})/2]}{\bar{q}^2-{\rm
sn}^2[((\varepsilon_2+1)^2-\n^2)^{1/2}(x-x^{(2)})/2]},\: {\mathcal{H}}_2>\n
\end{array} \right.
\end{equation}
The two parameters, $\varepsilon_2$ and $x^{(2)}$ are determined by
the boundary conditions at $x=x_1$. Here
$\varepsilon_2=(n_o^2+1+2{\mathcal{H}}_2)^{1/2}$,
$q=[(\varepsilon_2+n_o)^2-1]^{1/2}$ and
$\bar{q}=[(\varepsilon_2+n_o+1)/(\varepsilon_2+n_o-1)]^{1/2}$,
$k=[(n_o^2-(\varepsilon_2-1)^2)/4\varepsilon_2]^{1/2}$ and
$\bar{k}=[(\varepsilon_2-1)^2-n_o^2)/(\varepsilon_2+1)^2-n_o^2)]^{1/2}$ are
the moduli of the elliptic integrals of the first kind respectively for
the two cases. This solution is realized over about a quarter period,
where the electron density is positive and is given by
\beq \label{layer}
n(x)=\fr{2\sqrt{1+a^2}}{\n}\left({\mathcal{H}}_2+\fr{3}{2}\n
\sqrt{1+a^2}-a^2\right)
\eeq

The second boundary position of this layer can now be defined with a certain
arbitrariness, namely, it can be taken within the whole interval $x^* \leq x_2 \leq
x^{**}$, where $x^*$ is such that $n(x^*)=0$, while $x^{**}$ is defined by the
existence of a solution of the transcendental equation (\ref{trasc}). This problem
comes up for every layer, therefore the global solution, for a given incident
amplitude, in general may not be unique because the thickness of each plasma layer is
not uniquely fixed, the realization of each specific solution depending on the prehistory of
the process. This is also confirmed by our preliminary simulations based on a
hydrodynamics approach where, at fixed incident amplitude, the thickness of the
various electron layers in the quasi-stationary stage is quite different depending on
the turn-on shape of the incident radiation.  The appearance of non-unique solutions is typical of phenomena involving
nonlinear media (see, for instance, \cite{litvak}).

For the sake of concreteness, in what follows, we choose the next boundary position $x_2$
as the point where the electron density vanishes, i.e.
$n(x_2)=0$. In this case, the transcendental equation (\ref{trasc}) always has
nontrivial solutions because, if the condition given by Eq. (\ref{gprime}) is
satisfied for the semi-infinite layer, it will automatically be satisfied for all the
precedent electron layers, for the point where the electron density vanishes.\\
 As follows from (\ref{layer}), the field at this boundary is
such that
\begin{equation}
3n_{o}(1+a_{2}^2)^{1/2}=2(a_{2}^2 - {\mathcal{H}}_2)
\end{equation}
and, making use of the equation for the Hamiltonian (\ref{H02}), we can
calculate the first derivative of the field. Thus, for the next depletion
layer located within $x_3< x < x_2$, we know the boundary values of $a_2$
and $a^\prime_2$ that are required to generate a solution by using the same
procedure as before for $x_1< x < x_0$. This procedure is repeated for
every layer, $x_{i+1}<x<x_i$, up to the final layer and if
${\mathcal{H}}_i>\n$, the field solution has to be taken as in the second expression
of (\ref{sol1}). The last layer will be a depletion region $0 < x
< x_N$ where $x=0$ is the real plasma-vacuum boundary. At this
point the electric field due to charge separation must vanish. Integrating
Poisson's equation over the plasma interval, the plasma neutrality condition
gives
\beq
x_0=\frac{1}{n_o}\frac{a_Na^\prime_N}{\sqrt{1+a_N^2}}
\eeq
which defines the last free parameter. Thus, matching the solution for the
field to the vacuum solution at the electron layer boundary $x=x_N$ we can
relate the obtained plasma-field distribution to the incident
electromagnetic wave. In this way we can construct an exact stationary
solution for the anomalous penetration regime. An example of such a
solution is presented in Fig. 2. Notice that the maximum of the electron
density in a layer increases from layer to layer, reaching an absolute
maximum in the last layer nearest to the vacuum boundary while the width of
the layers become more and more narrow. This may easily be understood from
Fig. 1; higher lying trajectories have higher values of the integrals of
motion. It also means that at higher incident wave amplitudes there are
more layers, so that the penetration length will increase with increasing incident
intensity. These solutions can naturally be extended to cover the case of plasma slabs
with finite thickness, if the penetration length is smaller than the slab thickness.

The following consequence of the previous analysis should be emphasized: For a fixed amplitude the laser field
penetrates only a finite length into the overdense plasma. Consequently,
if the incident laser
pulse has a finite duration, the
electromagnetic energy deposited in the plasma will be reflected back into
the vacuum  after the laser field has vanished. Evidently the transient regime will be more complicated as
vacuum (depletion) regions surrounded by electron layers will show a
resonator-like behaviour, with the electromagnetic energy being excited by
the incident pulse. We expect this field structure to evolve and to be
slowly reflected back into the vacuum region.
% Simulations of laser-plasma interactions
%based on the hydrodynamics approach, where plasma kinetic effects do not shade or
%complicate the effects under analysis, confirm the scenario discussed above and will
%be presented elsewhere.

In conclusion, we have presented a new class of exact analytical stationary
solutions describing  a  new feature of the interaction between a
super-intense laser and an overdense plasma. This analysis shows how,
depending on the initial plasma density, the interaction might result in the
generation of a new plasma-field structure consisting of alternating
electron and "vacuum" regions, with the electromagnetic energy penetrating
into the overdense plasma over a finite length determined by the incident
intensity.

This work was supported in part by the Russian Fund for Fundamental
Research (grants No. 98-02-17015 and No. 98-02-17013). One of the authors (F.C.)
acknowledges support from the European Community under the contract ERBFMBICT972428.

\section*{Captions}

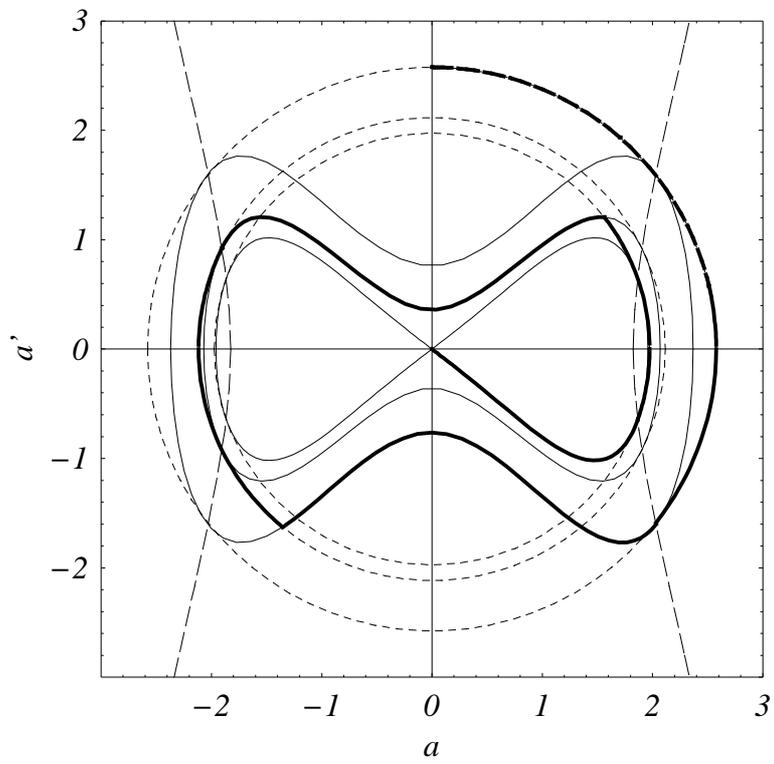
{\bf Fig. 1} Phase portrait of the system for $\n=1.6$ and homogeneous ion density,
following from Eqs. (\ref{int}) and (\ref{intvacuum}), see also Fig. 2. The dashed lines stand
for vacuum regions,
 the continuous lines stand for plasma regions, the actual trajectory is given by the
 thick line and it runs clockwise. The large dashed line denotes the
 regions where the electron density is negative.\\

\vspace{1 cm}

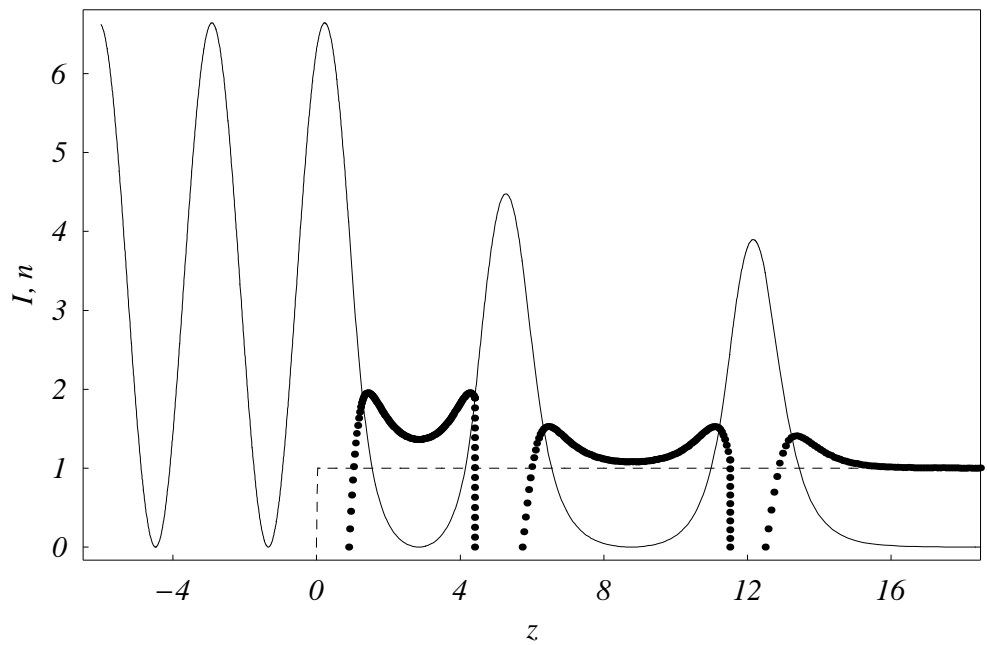
{\bf Fig. 2} The continuous line represents the plasma-field structures
in a semi-infinite plasma
initially occupying the region $x \geq 0$, for $\n=1.6$, the unperturbed electron density
being represented by the dashed line. The dotted line
represents the resulting electron density distribution.
All quantities are dimensionless.\\

\end{document}